\documentclass{emulateapj}

\usepackage{graphicx}

\usepackage{natbib}

\newcommand{\be}{\begin{equation}}
\newcommand{\ee}{\end{equation}}
\newcommand{\nn}{\mbox{} \nonumber \\ \mbox{} }
\newcommand{\ba}{\begin{eqnarray}}
\newcommand{\ea}{\end{eqnarray}}
\newcommand{\om}{\omega}
\newcommand{\Alfven}{ Alfv\'{e}n }

\newcommand\etal{\textit{et al.\ }}
\newcommand\eg{\textit{e.g.\ }}

\newcommand{\Chandra}{{\it Chandra\,}}

\begin{document}

\title{Dissipation in  intercluster plasma}
\author{Maxim Lyutikov}
\affiliation{Department of Physics, Purdue University, 525 Northwestern Avenue
West Lafayette, IN
47907 }

\begin{abstract}
We discuss dissipative processes in  strongly gyrotropic, nearly collisionless plasma in clusters of galaxies (ICM). First, we point out that  \cite{Braginsky} theory, which assumes that collisions are more frequent that the system's dynamical time scale,  is inapplicable to fast, sub-viscous ICM motion.  Most importantly,    the electron contribution to  collisional magneto-viscosity  dominates over that of  ions  for short-scale Alfvenic motions with wave length satisfying
$ l  \leq  {   \lambda \over \sqrt{\beta}} \left( {m_e\over m_p} \right)^{1/4}  \sim 1 {\rm  kpc}$ (where $\lambda $ is particle's mean free path, $\beta$ is the plasma pressure  parameter and $m_{e,p}$ are electron and proton masses). 
Thus, if a turbulent cascade develops in the ICM and propagates 
 down to  scales $\leq 1$ kpc, it  is damped collisionally 
not on ions, but on electrons.

Second, in  high beta plasma of ICM,  
small variations of the magnetic field strength, of relative value $\sim 1/\beta$, lead to development of anisotropic pressure instabilities (firehose, mirror and cyclotron). Unstable wave modes may    provide additional  resonant scattering of particles,  effectively keeping the plasma in a state of marginal stability. We show that  in this case the  dissipation rate of a   laminar, subsonic, incompressible flows 
scales as inverse of plasma beta parameter. We discuss application to the problem of ICM heating.
 \end{abstract}


\maketitle

\section{Introduction}

One of the key  problems in physics of intercluster medium (ICM)
is the absence of strong cooling flows at the centers of galaxy clusters
\citep[see, \eg,][for a review]{Peterson06}. 
It has been proposed that  heating of ICM by Active Galactic Nuclei (AGNs)
may be sufficient to offset the cooling \citep[\eg][]{begelman04}.
While the total energy budget of AGNs is, in principal, sufficient 
to offset the radiative cooling, details of how this is achieved are 
far from clear.

The observational confirmation of the AGN heating model comes from ubiquitous presence of AGN blown bubbles, identified 
by decreased X-ray emission in \Chandra and XMM maps \citep{McNamara}.
These bubbles expand and rise in the cluster
potentials transferring part of their
 energy to the  internal energy of ICM.
It has been suggested that this process  can be very efficient, so that
a large fraction of the power released by AGN 
ends up as internal energy of ICM. 

The 
high efficiency of energy dissipation is far from obvious. 
What is required is a distributed increase of the entropy of the gas, not just of the
internal or bulk energy
\citep[entropy floor problem][]{lloyd00}.
The main problem is that these  AGNs blown bubbles expand, typically, 
{\it subsonically}, as is indicated by the general absence of  shock signatures
ahead of the  bubbles. 
 In laminar flows at small Reynolds numbers,  $Re \leq Re_{crit} \sim 10-100$, dissipation efficiency is $\propto 1/Re$. For  $Re \sim 100$, such a  low efficiency   puts unreasonable  
demands on AGN luminosity.


\section{Collisional dissipation in gyrotropic plasma: gyrorelaxational heating }
\label{2}

Ion 
Larmor radii in ICM,  $r_L \sim 10^8$ cm, 
is  some fifteen orders of magnitude  smaller than the 
system size, $L\sim $ hundreds of kpc,   and Coulomb
 mean free path,   $\lambda \sim 10-30 $  kpc,   for a typical density $n\sim 10^{-3}$ cc, magnetic fields $\sim 1-10 \mu$G   and
 temperatures in the keV range \citep[\eg][]{carillitaylor02}. Thus ICM is  weakly collisional, $r_L \ll \lambda$. 
In addition, it is weakly magnetized, in a sense that magnetic fields energy
is smaller than plasma pressure, $\beta =  8 \pi P / B^2 \geq 1 $.
Below we will refer to this regime as a strongly 
gyrotropic  plasma.

Dissipation in a strongly gyrotropic plasma proceeds in a qualitatively different way from the isotropic case, as is exemplified by so called 
gyrorelaxational heating.
If in an initially pressure-isotropic plasma  the absolute value of 
magnetic field oscillates with a frequency $\om$ and relative amplitude
$\delta = \delta B /B_0$, then
the dissipation rate $\alpha$  (so that energy of a particle ${\cal E} $ changes according to $
{d {\cal E}  / dt}= \alpha {\cal  E}$) in a cycle is \citep[eg][]{borovsky}
\be
\alpha \approx {\om^2 \nu_c \over {9\over 4} \nu_c^2 +\om^2}  {\delta^2 \over 6}
\label{alpha}
\ee
where $\nu_c $ is collision frequency.  Dissipation of energy occurs both due to electron and ion collisions, so that $\alpha=\alpha_e+\alpha_i$, calculated with corresponding collision frequencies 
$\nu_e$ and $\nu_i$. 
 In the high collision frequency regime,
$\nu_c \gg \om$,  Eq. (\ref{alpha}) approximates Braginsky's result, $\alpha \propto 1/ \nu_c$ \cite{Braginsky}.
Since ions have smaller collision frequency, dissipation in this limit is dominated by ions.
On the other hand, {\it for rare collisions,
$\nu_c \ll \om$, dissipation rate is proportional to collision frequency, $\alpha \propto  \nu_c$, and is thus dominated by  electrons} for  $t_e \om > {3\over 2} (m_e/m_i)^{1/4}\approx 0.2$


Consider sub-viscous turbulent  motion of ICM occurring on scale $l$ smaller than mean free path $\lambda$ and mediated by \Alfven waves, so that a typical wave frequency is $\om = V_A/l \sim 
c_s/ (\sqrt{\beta} l)$. Then for waves satisfying
$t_e \om >  {3\over 2} (m_e/m_i)^{1/4}$, or for
\be
l  \leq  {   \lambda \over \sqrt{\beta}} \left( {m_e\over m_p} \right)^{1/4}  \sim 1 {\rm  kpc}.
\ee
electron viscosity  dominates over ion viscosity.
For numerical estimates we assumed  $T=10^8$ K, $n=10^{-3}$ cc, $B=5 \, \mu$ G, so that $\beta \sim 10$  and mean free path $\lambda = 23 $ kpc.

 Thus, if a turbulent cascade develops in the ICM and propagates 
 down to   scales $\leq 1$ kpc, it  is damped  collisionally 
not on ions, but on electrons. 
Thus, {\it  \cite{Braginsky} theory, which assumes frequent collisions, $t_{coll} \om \gg 1$,  is inapplicable to fast, sub-viscous ICM motion. }

\section{Heating in a bound anisotropy model}
\label{3.1}
Besides binary collisions, plasma can be 
 heated through development of electromagnetic turbulence which resonantly scatters particles and, thus, provides an additional dissipation mechanism. 
  In this Section we  describe such mechanism of dissipation  through development  anisotropic plasma instabilities.

\subsection{Viscosity due to binary collisions in a gyrotropic plasma}

When Coulomb collision frequency $\nu _c$  is much smaller than cyclotron frequency, $\om_B / \nu_c \gg 1$,
plasma viscosity is strongly
 anisotropic, determined by seven coefficients \citep{Braginsky}. 
 In the limit $\om_B \rightarrow \infty$ and slow changes of magnetic field, $\om  \ll \nu_c$, the only remaining coefficient
 is $\eta_0$, responsible for the viscosity along the field lines.
 In this case the viscose
 stress tensor becomes \citep{LLX} 
\be
\sigma_{ij} = \eta_0 
\left( 3 b_i b_j -\delta_{ij} \right) \left( b_l b_k \partial_l v_k - {1\over 3}{\rm  div} {\bf v} \right)
\label{sigma}
\ee
where $b_i$ is a unit vector along the local magnetic field,  $\eta_0= p/\nu_c$, $p=(P_\parallel+ 2P_\perp)/3$ is total pressure. Below we concentrate on the incompressible limit, $ {\rm div} {\bf v}=0$, which  eliminates reversible compressional heating. For
 incompressible plasma without conductivity, using   Eq. (\ref{sigma}), the volumetric dissipation and entropy generation  rates due to viscosity are  \citep{LLVI} 
\be
\rho {d E \over dt}= \rho T {dS \over dt}
=\sigma_{ij} \partial_i v_j =  3 \eta_0 \left( {\bf b}  \cdot   ( {\bf b} \nabla) {\bf v}  \right) ^2
\ee
Using  induction equation,
$
{d  {\bf B}/ dt} = ({\bf B} \nabla) {\bf v},
$ 
  the entropy generation rate can be related to the rate of change of magnetic field \citep{scheko06}:
\be
\rho {d E \over dt}=  3 \eta_0  \left( { 1\over B} {d   B \over dt} \right)^2
\label{Q}
\ee
Dissipated power of a  gyrotropic fluid
 is  solely due to changing magnetic field, which is  very different from the isotropic case. 
This result can also be verified if we note that  in a  
  gyrotropic plasma the entropy is  
 $S \propto(1/2) \ln P_\perp P_\parallel ^2$ (assuming constant density).
The 
entropy production is then 
\be
{d S \over dt}=
{1 \over 3} { (P_\perp- P_\parallel)^2 \over P_\perp P_\parallel} \nu
\label{DS}
\ee
For binary collisions using $P_\perp- P_\parallel= 3 \eta_0 d_t \ln B$ (Eq. (\ref{31})), this gives
\be
{d S \over dt}= 3 (d_t \ln B)^2 \eta_0^2 {\nu \over P_\perp P_\parallel}
\approx 3 {(d_t \ln B)^2 \over \nu}
\label{DS1}
\ee
consistent with (\ref{Q}).

The differences between the dissipation rates calculated using isotropic and anisotropic viscosities 
can be dramatic. For example, for spherical expansion of a
bubble into incompressible fluid, in absence of magnetic field, the dissipated power is zero (flow field is irrotational). 
Introduction of  a weak (in a sense that $\beta \gg 1$) magnetic field
changes this picture completely.
In a kinematic approximation (neglecting its dynamical effects, so that
 field lines are just advected with the flow satisfying frozen-in condition) expansion of a bubble into a constant magnetic field creates magnetic fields
\be
B_\theta = {  \sin \theta \over (1- \xi^{-3})^{1/3}} B_0
, \,\,\,
B_r = - \cos \theta (1- \xi^{-3})^{1/3} B_0
\label{BB}
\ee
where  $\xi =r/R(t)> 1$ and $B_\theta$ and $B_r$ are component of magnetic field in a spherical 
system of coordinates aligned with the  initial direction of magnetic field.
Though tangential component of magnetic field diverges on the contact $\xi=1$ (magnetic draping effect), the 
 increase in B-field energy over the initial homogeneous field is finite,
$
 ({1 / 9}) B_0^2 R^3
$ and the total heating rate is $
d_t E= 3 \eta_0 R^3 \int d^3  \xi \left( {d _t \ln B} \right)^2 =  9.54 \eta_0  R^3 (d \ln R / dt)^2$.


This  example illustrates an important point: even a weak
magnetic field may considerably affect plasma dissipative properties.
Inverse situation, when a dissipative flow with isotropic viscosity becomes  non-dissipative in the strongly gyrotropic limit, is also possible. The example is a longitudinal shear, when magnetic field is directed along velocity. In the absence of cross-field viscosity there is no dissipation.

\subsection{Anisotropic pressure instabilities}

In collisionless plasma, particles in magnetic fields tend to conserve their
adiabatic invariants \citep{Chew}.
In case of rare collisions the equations describing evolution of pressures becomes
 \cite[eg][]{hollweg85}
 \ba && 
{d \ln P_\perp/B \over dt} = {\nu \over 3} {P_\parallel-P_\perp \over P_\perp}
\nn &&
{d \ln P_\parallel B^2 \over dt} = -{ 2 \nu \over 3} {P_\parallel-P_\perp \over P_\parallel}
\label{Eq}
\ea
where $P_\perp$ and $P_\parallel$ are pressure across and along magnetic field.

In a $\beta \gg 1$ plasma, the
development of pressure anisotropy 
 may lead to  firehose, mirror and  ion cyclotron    instabilities when the
 following conditions are satisfied
\be 
\beta_\parallel - \beta_\perp> 2,  \mbox{\,firehose}, \,
{\beta_\perp \over\beta_\parallel}> 1+ {1 \over \beta_\perp}
\mbox{\,mirror}
\label{inst}
\ee and 
${\beta_\perp / \beta_\parallel}> 1+ {k / \beta_\parallel^m}
\mbox{\,cyclotron}$,
where $\beta_\parallel = 8 \pi P_\parallel/B^2$,
$\beta_\perp = 8 \pi P_\perp/B^2$,
$0.35 \leq k  \leq 0.65$ and $0.4  \leq m  \leq 0.42$ \cite{Gary94}.
Cyclotron instability has  growth rate larger than the  mirror instability  
for  $\beta \leq 6$ and $p_\perp > p_\parallel$.
 If initially plasma  pressure is isotropic,  firehose and mirror instability occur when 
$
{\delta B  / B } =  -  {2 /( 3 \beta_0)} \mbox{(firehose)}, \,
{\delta B / B } = + {1 /( 3 \beta_0)}  \mbox{(mirror)}
\label{Binst}
$
and similar expression for ion cyclotron    instability \citep{Gary94}; for clarity we do not consider it here.

The instabilities' increment is maximal at the cyclotron frequency, which is very fast compare
to any dynamical time. Further change of magnetic field, beyond the limits (\ref{Binst}),
 will be accompanied by development of  instabilities
 which  will lead to
increased scattering rate, either due to quasi-linear diffusion or a fully
 developed turbulence. As a result,
   the system dissipates quickly any free energy in excess of instability threshold and relaxes
   to the marginally  stable state. We expect that the system remains at threshold
   of instability.
   

\subsubsection{Binary collisions in sub-critical regime}

 Binary collisions decrease a level of anisotropy and may stabilize plasma. 
 Redefining
pressures  $P_\perp$ and $P_\parallel $  in terms of total pressure $p$
(a trace of the pressure tensor) and pressure disbalance $( P_\perp - P_\parallel)/p = \Delta$,
$P_\parallel =  p - {2 \over 3}  \Delta p,\,\,\,
P_\perp =p+{ 1 \over 3} \Delta p
$,
we find 
\ba &&
2 p  \Delta {d B \over dt}  = 3 B {d p \over dt}
\nn &&
{d  \Delta \over dt}  + \nu \Delta  - {9 - 3  \Delta - 2  \Delta^2 \over 3}  {d \ln B \over dt} =0
\label{dd}
\ea
In a $\beta \gg 1$ plasma 
   at the moment of instability $\Delta $ is small,  $|\Delta| \ll 1$. 
For  slow changes $d/dt \ll \nu$  this  gives
\be
\nu \Delta= 3 {d \ln  B \over dt}
\label{31}
\ee
This implies that  for development of instabilities  the dynamical time
   $t_{dyn} \sim 1/ d_t \ln B$ should be relatively short, 
    $
  {  t_{dyn}  \nu_c }  \leq  \beta
      $.
This condition is satisfied by most scales of interest in  ICM plasma.

\subsection{Dissipation at marginal stability}

As we argued in the previous section, changing magnetic field will lead to development of  instabilities
that will keep the plasma anisotropy at the critical values given by Eqs.
(\ref{inst}).
Eqs. (\ref{Eq})
and the condition (\ref{inst}) may be regarded as defining an effective
scattering rates 
\be
\nu_{eff, firehose} = {3\over 2} \beta d_t \ln B,\,\,
\nu_{eff,mirror} = {3} \beta d_t \ln B
\label{nueff}
\ee

%
At a critically balanced case,  the entropy generation rate Eq. (\ref{DS}) 
 \be
d _t S  
\approx 
{2 \over \beta} d_t \ln B
 \times \left(
\begin{array}{l}
1
\\
{1 \over 2}
\end{array}
\right)
\ee
for the firehose and mirror regimes.

We have arrive at an important result related to efficiency of dissipation:
{\it in a gyrotropic plasma efficiency of  dissipation is determined not by Reynolds number, but by 
the plasma beta parameter}. Typical dissipation time scale is  $\beta$ 
times dynamical time, not $Re$ times dynamical time.



The role of effective
collisions in  energy dissipation in a marginally stable regime is, in some sense, opposite
to  the role of binary collisions in a sub-critical regime. The entropy production rate and corresponding  volumetric dissipated power, Eq. (\ref{DS}), are proportional to pressure anisotropy and collision frequency, 
 $\propto \nu (\Delta P)^2 $, where $\Delta P$ is the difference in parallel and transverse pressures. If the pressure disbalance is due to binary collisions,  then $\Delta P \propto 1/\nu$ 
so that the dissipation rate is $ \propto 1/\nu$ \citep{Braginsky}. Thus, before the instabilities are reached,   
 increasing collision rate leads to decreasing
dissipation.  On the other hand, for marginally
stable case $\Delta P \sim $ constant, so that dissipated power is {\it proportional} to the effective
collision rate, Eq. (\ref{DS1}).

\subsection{Damping of waves at marginal stability}

For \Alfven waves,  perturbations of magnetic field are orthogonal to the initial magnetic field, so that
variations of the absolute value of the field are second order in amplitude. 
For large enough amplitude, satisfying condition $(\delta B/B_0)^2 \equiv \delta^2 > 1/\beta$, this creates conditions
favorable for mirror instability.
The 
entropy production rate over the period is
\be
{dS \over dt} ={  4 \over  \beta}{ \delta ^2  \over 1+ \delta^2} \om \left( { 2 \arccos {1\over  \beta \delta^2} \over \pi} \right)
\ee
where the term in parenthesis takes into account 
 phases when the amplitude of  fluctuations satisfies the mirror instability criterion. With Braginsky viscosity, the collisional damping of \Alfven waves is a non-linear effect as well, but it has a much steeper dependence of wave amplitude and frequency. From   (\ref{Q})  we find
\be
{dS \over dt} ={3 \delta^4 \om^2 \over (1+\delta^2)^2 \nu_c}
\ee
For comparison, in isotropic MHD \Alfven waves  are damped at  a rate
 (\cite{LLVII})
$
{dS / dt}={ \delta^2  \om^2 / \nu_c}
$.

\section{Discussion}

Our approach follows a long established procedure   of  marginal stability
\cite{KennelPetschek66,manheimer,Gary94,Denton94}, when the instability threshold becomes the limiting value of anisotropy.  
In particular, \cite{Quest,Gary98} applied a  bounded anisotropy model  to the measurements of parallel and perpendicular temperatures in the solar bow shock region near the Earth
magnetosphere. 
 It was  found that  an initial rapid growth of unstable waves indeed brings the system back to  approximate marginal stability.

What is the relation of the marginal stability condition and the conventional quasilinear and turbulence theories? According to  \cite{manheimer},  both predict some level of turbulent fluctuations. Marginal stability approach is applicable if the level of those fluctuations is smaller than the one calculated from non-linear theory. This, typically, happens when the driver of the instability (in our case a large scale motion of ICM plasma) is not strong. Assessing whether this is satisfied in case of ICM plasma requires full scale calculations of non-linear turbulence levels, a prohibitively complicated task  given the uncertainties in both plasma microphysics and  details of ICM  plasma motions. 

The most important effect that was not taken into account in the present work is thermal conduction. The double-adiabatic equations
are valid only when heat flux 
along magnetic field lines
can be neglected. 
This is the main reason why the theory may  fail \cite[\eg, the notorious results of][]{Kulsrud}.
Neglect of heat flux requires  that phase velocity of the perturbations be much larger than speed of heat carriers, electrons: $\left({\om \over k}  \right)^2 \sim V^2 \gg v_{T,e}^2$. This condition may be broken in ICM, especially outside of cluster cores. 
On the other hand, enhanced scattering rate suppresses conductivity \citep{LevinsonEichler}. 
The conduction coefficient 
is 
$
\kappa \sim n_e v_{T,e}^2 /\nu_{eff} 
$ (assuming that saturate conductivity regime  \citep{cowie} is not reached).
The effective scattering frequency due to development of electromagnetic instabilities Eq. (\ref{nueff})
may be higher than  binary collision rate,  so that the
 conduction coefficient 
 will be smaller,
$
\kappa \sim n_e {  v_{T,e}^2 L/( \beta V)} 
$.  Increased scattering will also  inhibit the onset of  saturated regime.

There is a number of challenges that heating models  should overcome. Primarily, 
the heating must be
both widely distributed and gentle. It is hardly achievable with shocks, which provide very concentrated heating  at the shock location, deposit most of the energy in the core and generally contradict the observational  absence of shock signatures.
 This, combined with low heat conductivity in the cores,  leads to plasma overheating and creation of   inverted entropy gradients, contrary to observations \citep[\eg][]{voit01}.  
 
 The heating in the bounded anisotropy model may  be distributed. Consider a cluster with a typical  density profile
 $\rho \propto 1/r$. Then if bremsstrahlung dominates over line emission, the cooling rate  is $\propto r^{-2}$ (for nearly constant temperature in the cores). 
Since an energy flux from central source scales as $\propto r^{-2}$ as well, this implies that a heating rate 
should be independent of a radius, and thus independent of the local plasma properties. Collisional dissipation clearly cannot produce this. On the other hand, if $\beta$ is nearly constant, the heating rate will be nearly independent of radius.
Thus, at least in principle, heating and cooling can be balanced in the bound anisotropy model. 

One of the main drawbacks of many simulations of ICM is that they use isotropic Spitzer  viscosity.  Examples in \S \ref{3.1} show that this can produce (at least locally) drastically incorrect results, which may  either overestimate or underestimate the real collisional magnetoviscosity (we are not aware of any ICM-related simulations with anisotropic viscosity \citep[see, though,][]{Sharma}). As for the value of the coefficient of viscosity, we argued that for binary collision it generally depends on  electron and ion temperatures and  dynamical times scales, while  in case of marginal stability it is  actually  unrelated to the Spitzer   value. 
Parametrization with respect to  Spitzer may be  useful, but  we should not put too much physical emphasis on it.


 \end{document}